\documentclass[aps,prl,twocolumn,superscriptaddress,fleqn]{revtex4-1}

\usepackage{lipsum}
\usepackage{graphicx}
\usepackage{amsmath, amsfonts, amssymb, amsbsy}
\usepackage[usenames,dvipsnames]{xcolor}
\usepackage{marvosym}
\usepackage{sidecap}
\usepackage{wrapfig}
\usepackage[utf8]{inputenc}

\newcommand{\beq} {\begin{equation}}
\newcommand{\eeq} {\end{equation}}

\newcommand{\Sei} {Se$_{\rm in}$} 
\newcommand{\Seo} {Se$_{\rm out}$} 
\newcommand{\bise} {Bi$_2$Se$_3$}
\newcommand{\cubise} {Cu$_{0.15}$Bi$_2$Se$_3$}
\newcommand{\bite} {Bi$_2$Te$_3$}

\newcommand{\super} {\textsuperscript} 
\begin{document}


\title{$^{77}$Se nuclear magnetic resonance of topological insulator Bi$_2$Se$_3$}


\author{Nataliya M. Georgieva}
\affiliation{Faculty of Physics and Earth Sciences, University of Leipzig, Linn\'{e}strasse 5, 04103 Leipzig, Germany}
\author{Damian Rybicki}
\affiliation{Faculty of Physics and Earth Sciences, University of Leipzig, Linn\'{e}strasse 5, 04103 Leipzig, Germany}
\affiliation{Faculty of Physics and Applied Computer Science, AGH University of Science and Technology, Department of Solid State Physics, al. A. Mickiewicza 30, 30-059 Krakow, Poland}
\author{Robin Guehne}
\affiliation{Faculty of Physics and Earth Sciences, University of Leipzig, Linn\'{e}strasse 5, 04103 Leipzig, Germany}
\author{Grant V. M. Williams}
\affiliation{School of Chemical and Physical Sciences, Victoria University of Wellington, PO Box 600, Wellington 6140, New Zealand}
\author{Shen V. Chong}
\affiliation{Robinson Research Institute, Victoria University of
Wellington, PO Box 33436, Lower Hutt 5046, New Zealand}
\author{Kazuo Kadowaki}
\affiliation{Division of Materials Science, Faculty of Pure and Applied Sciences, University of Tsukuba, 1-1-1, Tennodai, Tsukuba, Ibaraki 305-8573, Japan}
\author{Ion Garate}
\affiliation{D\'{e}partement de Physique and Regroupement Qu\'{e}b\'{e}cois sur les Matériaux de Pointe, Universit\'{e} de Sherbrooke, Sherbrooke, Qu\'{e}bec, Canada J1K 2R1}
\author{J\"urgen Haase}
\affiliation{Faculty of Physics and Earth Sciences, University of Leipzig, Linn\'{e}strasse 5, 04103 Leipzig, Germany}

\date{\today}

\begin{abstract}
Topological insulators (TIs) constitute a new class of materials with an energy gap in the bulk and peculiar metallic states on the surface. 
To date, most experiments have focused on probing the surface electronic structure of these materials.
Here, we report on new and potentially interesting features resulting from the {\em bulk} electronic structure. 
Our findings are based on a comprehensive nuclear magnetic resonance (NMR) study of $^{77}$Se on Bi$_2$Se$_3$~and Cu$_{0.15}$Bi$_2$Se$_3$ single crystals. 
First, we find two resonance lines and show that they originate from the two inequivalent Se lattice sites.
Second, we observe unusual field-independent linewidths, and attribute them to an unexpectedly strong internuclear coupling mediated by bulk electrons. 
These results call for a revision of earlier NMR studies and add insight into the bulk electronic properties of TIs.
\end{abstract}




\maketitle

\begin{figure}[t]
\includegraphics[width=0.45\textwidth]{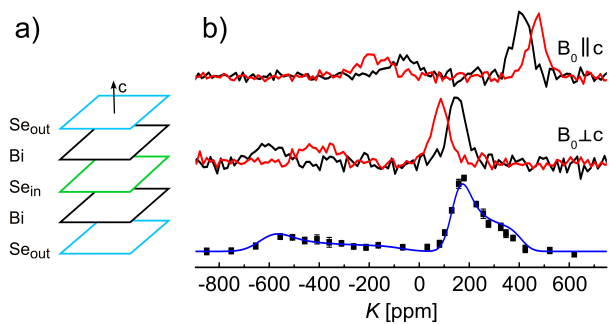}
\caption{\label{fig:spec}(color online). a) Sketch of the quintuple layer. b) $^{77}$Se NMR spectra at $B_0$ = 17.6 T and room temperature of Bi$_{2}$Se$_{3}$ (black) and \cubise\, (red) single crystals for two crystal orientations (top two), and of \bise\, powder (squares) with simulation based on single crystal data (solid blue line). Shifts ($K$) are given with respect to (CH$_{3}$)$_{2}$Se.}
\end{figure}

The discovery of topological phases of matter in three dimensions has sparked great interest in the scientific community~\cite{Hasan2010, Qi2011}. 
Three-dimensional topological insulators (TIs)  were predicted \cite{Fu2007, Moore2007, Roy2009, Zhang2009} and subsequently confirmed \cite{Hsieh2008, Xia2009, Chen2009, Hsieh2009} in spin-orbit coupled systems with inverted band structures.
Among these, \bise\, has emerged as a model system due to its simple surface states and due to the relative ease with which it can be synthesized in the form of large single crystals
.  
The crystal structure of \bise\, consists of stacked, van der Waals bonded quintuple layers (QL) of five atomic sheets each, with the $c$-axis normal to the layers, cf.~Fig.~\ref{fig:spec}(a).  
Each QL contains two equivalent ''outer'' Se atoms (\Seo), two equivalent Bi atoms,  and another ''inner'' Se atom (\Sei) located at the center of inversion \cite{Zhang2009}.
In spite of its energy gap, the bulk of \bise \,is conducting due to self-doping with electrons from Se vacancies, with  carrier concentrations ($n$) ranging from $2\times10^{17}$ to $2\times10^{19}$ cm$^{-3}$ \cite{Hor2009, Analytis2010, Wang2010, Nisson2013}, which can be increased, e.g., by intercalation of Cu~\cite{Hor2010, Kriener2011, Bay2012}.

NMR is a powerful probe of chemical and electronic material properties, but it is unclear what it can contribute to the understanding of the protected surface states or to any special bulk properties of TIs (NMR has been proposed to probe the pairing symmetry of topological superconductors \cite{Zocher2013}). 
So far, there have been few NMR studies of TIs: $^{209}$Bi NMR of \bise\, single crystals and powders \cite{Young2012, Nisson2013, Nisson2014, Mukhopadhyay2015}, $^{125}$Te NMR of \bite\, \cite{Taylor2012, Koumoulis2013, Koumoulis2014, Podorozhkin2015}, and $^{77}$Se NMR of \bise\, powder \cite{Taylor2012}. 
While all of these confirm bulk conductivity qualitatively from fast longitudinal nuclear relaxation ($1/T_1$), they leave many questions unanswered. 
For example, although the two inequivalent Se (nuclear spin $I=1/2$) or Te ($I=1/2$) sites should give rise to different NMR signals, they have not been found or discussed, while signals from surface states have been reported \cite{Koumoulis2013, Podorozhkin2015}. Even where reported, special lineshapes or spin echo behaviors are not understood, e.g., for $^{209}$Bi NMR ($I=9/2$) ~\cite{Young2012, Nisson2013, Nisson2014}, pointing to unusual electronic properties. Therefore, understanding the NMR of TIs may open up the possibility of a more detailed comprehension of these materials, including the characterization of surface states.

\begin{figure}[t]
\includegraphics[width=0.4\textwidth]{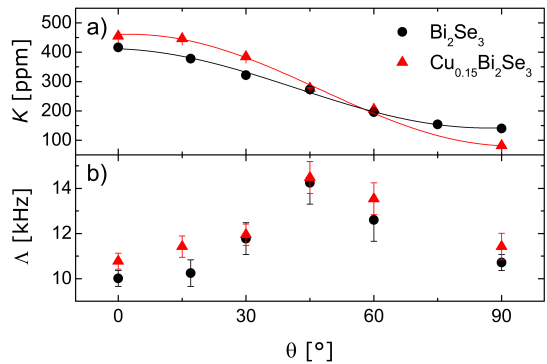}
\caption{\label{fig:ang}(color online). Angular dependences of shifts $K$ (a) and linewidths $\Lambda=\sqrt{\ln{4}}/(\pi T_{2G}^*)$ (b) of \Seo\, in \bise\, and \cubise\, single crystals at 17.6 T. The solid lines in (a) are fits to $K=K_{\rm iso} +\Delta K (3\cos^2\theta -1)/2$.}
\end{figure}

Herein, we report on mostly $^{77}$Se NMR studies of \bise \,and \cubise.
We identify two bulk Se signals that we prove to be due to Se$_{\rm in}$ and Se$_{\rm out}$. 
We find NMR shifts, relaxation, linewidths, and spin echo decays to be quite different at these two sites.  
We discover a strong indirect internuclear coupling that is mediated by the bulk electrons and that is responsible for the unusual linewidths and echo decays of $^{77}$Se (probably also for $^{209}$Bi NMR).
Our results give new insight into the electronic properties of TIs, call for a revision of conclusions from earlier NMR studies, and help lay a foundation for the 
characterization of surface states with NMR.
 
\begin{figure}[b]
\includegraphics[width=0.9\columnwidth]{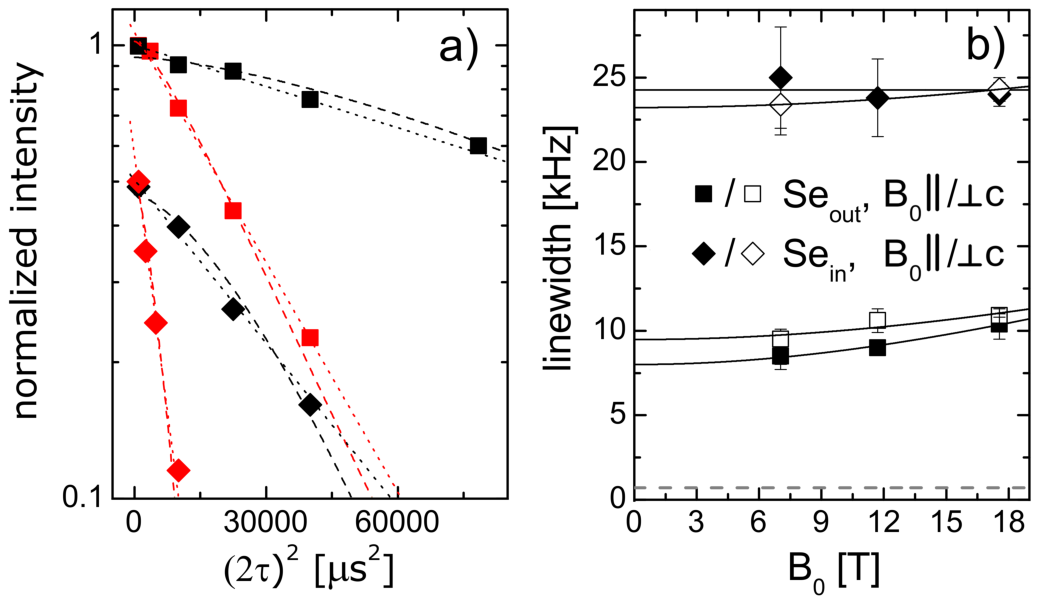}
\caption{\label{fig:DecayLWvsB}(color online). a) \bise\,\! (black) and \cubise\,\! (red), spin echo decays of \Seo\,\! (squares) and \Sei\,\! (diamonds) at 17.6 T for $B_0\!\parallel\!c$. Dotted lines are Gaussian fits proportional to $\exp\left(-(2\tau)^2/2T_{2G}^2\right)$. Dashed lines are Recchia fits (see text) with second moments taken from the field-independent linewidths in \bise\, from (b). b) \bise, linewidths ($\Lambda$) of \Seo\,\! and \Sei\,\! at different magnetic fields $B_0$. Solid lines are fits to ${\Lambda}^2=\Lambda_0^2+[b B_0]^2$, the dashed line represents the maximal linewidth expected from magnetic dipole interaction (see text).
}
\end{figure}

\begin{table*}
\begin{tabular}{lcccccccccc}{
\color{white}\bise}&&\multicolumn {4}{c}{\bf\bise}&&\multicolumn{4}{c}{\bf\cubise}\\
\hline
&{\color{white}1}& ~$K$ [ppm]~~ & ~~$T_1$ [s]~~ & ~~$\Lambda$ [kHz]~~ & ~$T_{\rm 2G}$ [$\mu$s]~&{\color{white}111}& ~$K$ [ppm]~~ & ~~$T_1$ [s]~~ & ~~$\Lambda$ [kHz]~~ & ~$T_{\rm 2G}$ [$\mu$s]~\\
\hline
$B_0\,\parallel\,c$, \Seo &&  {\color{white}-}410$\pm${\color{white}1}7 & 4.7$\pm$0.3 & 10.1$\pm$0.4 &  260$\pm$20&&
{\color{white}-}475$\pm${\color{white}1}5 & 0.40$\pm$0.03& 10.8$\pm$0.4 &  113$\pm$4\\ 
$B_0\,\parallel\,c$, \Sei &&  {\color{white}2}-63$\pm${\color{white}1}7 & 1.4$\pm$0.3 & 24.0$\pm$0.7 & 134$\pm${\color{white}1}2&& 
-175$\pm$10 & 0.13$\pm$0.01& 20.6$\pm$1.2 & {\color{white}1}54$\pm$2\\
$B_0\perp c$, \Seo && {\color{white}-}144$\pm${\color{white}1}7 & 3.0$\pm$0.3 & 10.9$\pm$0.1 &  215$\pm$10&&
{\color{white}1-}86$\pm${\color{white}1}5 & 0.21$\pm$0.02& 12.4$\pm$1.2 &{\color{white}1}99$\pm$5\\   
$B_0\perp c$, \Sei &&  -605$\pm$12 & 2.0$\pm$0.3 & 24.3$\pm$0.7 &  {\color{white}2}95$\pm$10&&
-400$\pm$10 & 0.24$\pm$0.03& 24.7$\pm$3.5 &  {\color{white}2}35$\pm$5\\ 
\end{tabular}
\caption{Measured total shift $K$, spin-lattice relaxation time $T_1$, linewidth $\Lambda$ and spin echo decay time $T_{2G}$ in \bise\, and \cubise\, at ambient conditions and 17.6 T.}
\label{tab:data}
\end{table*}

Single crystals of Bi$_2$Se$_3$ and \cubise\, have been grown as described in Ref.~\cite{Das2011}. 
Granular Bi$_2$Se$_3$ from Sigma Aldrich was ground to a fine powder prior to experiments. 
$^{77}$Se NMR was studied at B$_0$ = 7, 11.7, and 17.6 Tesla with standard wide-bore NMR magnets, home-built probes, and commercial consoles. 
The signals were acquired with spin echo pulse sequences ($\pi/2 - \tau - \pi$) with typical $\pi/2$ pulse lengths of 5 to 7 $\mu$s. 
$^{77}$Se has a low natural abundance ($7.63\%$); accordingly, even at the highest field about 2000 scans were necessary to obtain a sufficient signal-to-noise ratio. 
We have found the radio frequency (RF) tank circuits' quality factors ($Q=\omega L/r$) determined by losses from single crystals with changes of the apparent series resistance ($r$) rather than the inductance ($L$) \footnote{We have verified that no sample heating occurs during our measurements, despite the sample being the primary heat sink for RF power.}.
Typical quality factors were $Q\approx 30$.
We also estimated the absolute NMR intensities for the single crystal samples, i.e., the number of observed nuclei.
For calibration we used H$_2$SeO$_3$ powder with the same RF coil, and we corrected for differences in the quality factor and spin echo decay ($1/T_{\rm 2G}$). We find that at least 25\% of the $^{77}$Se nuclei  contribute to the signal. This corresponds to an RF penetration depth of at least 
80 $\mu$m in \cubise, in agreement with resistivity data \cite{Hor2010} that give 100 $\mu$m at 200 K for a Cu$_{0.12}$\bise\, sample with $n=2\times 10^{20}$ cm$^{-3}$. 
 
Typical \super{77}Se~NMR spectra are shown in the upper part of Fig.~\ref{fig:spec}(b). 
We identify two resonance lines, a narrower and a wider signal, with an intensity ratio of about 2:1. 
The spectrum of the Cu doped sample is similar to that of \bise, except for differences in the shifts. 
All signals are Gaussian functions of time, i.e., with the form $\exp\{-t^2/2T_{2G}^{*2}\}$.
The $T_1$ for both signals is a few seconds  in \bise, but an order of magnitude smaller in the Cu doped sample, see~Tab.~\ref{tab:data}. 
The powder spectrum of \bise\, in the bottom of Fig.~\ref{fig:spec}(b) consists of two regions (with an intensity ratio of about 2:1), in agreement with what we calculate from our single crystal data. 

Figure~\ref{fig:ang} depicts the dependence of the shifts and linewidths on the polar angle $\theta$ (between the crystal $c\,$-axis and the magnetic field $B_0$) for the narrow lines. 
These findings are in agreement with the crystal structure,
however, we estimate $^{77}$Se NMR linewidths from magnetic dipole interaction~\cite{Abragam1961} to be 0.7 kHz (1.3 kHz) for  \Seo\,  and 0.6 kHz (1.6 kHz) for \Sei,  with ${B_0\parallel c}$  (${B_0\perp c}$), while the experimental widths are an order of magnitude larger and with a much weaker angular dependence, cf. Tab.~\ref{tab:data}. 

Typical spin echo decays are shown in~Fig.~\ref{fig:DecayLWvsB}(a). 
The Gaussian decay constants ($T_{\rm 2G}$) range between 95 and 260 $\mu$s for \bise\, and are all shortened in \cubise\, by about a factor 2.4, cf.~Tab.~\ref{tab:data}. 
We estimate decay constants between 2300 and 4800 $\mu$s from homonuclear dipolar coupling between similar $^{77}$Se nuclei. 

Given the discrepancies between measured and expected linewidths as well as echo decays, we have investigated the magnetic field dependence of the linewidths. 
The results are shown in Fig.~\ref{fig:DecayLWvsB}(b). 
Surprisingly, we find large field-\textit{independent} linewidths, i.e., about 9 kHz for \Seo\, and 24 kHz for \Sei, while the field-dependent linewidths are less than $\sim$50 ppm (0.4 kHz/T).

We will now discuss our observations. 
Clearly, the two $^{77}$Se NMR signals originate from the outer (\Seo) and inner (\Sei) sites of the QL, because (i) an intensity ratio of 2:1 is expected from stoichiometry; (ii) we observe the bulk of all samples, and (iii) the powder spectrum with much higher surface area is in agreement with single crystal data. 
The powder spectrum reported in Ref.~\cite{Taylor2012} could not distinguish the two signals due to the large field-independent linewidths and their low magnetic field (7.05~T).
In the supplement of Ref.~\cite{Mukhopadhyay2015}, a single crystal $^{77}$Se NMR spectrum consisting of one line with a field-independent $\sim$12 kHz linewidth is reported. This suggests the authors observed only \Seo, while \Sei\, was missing, probably due to noise.
Since \bite\, is structurally similar to \bise, we anticipate the existence of two bulk Te signals therein. Only a single bulk $^{125}$Te NMR signal has been resolved so far \cite{Taylor2012,Koumoulis2013,Podorozhkin2015}, while a second signal has been ascribed to surface nuclei \cite{Koumoulis2013,Podorozhkin2015}.
In view of our results, these interpretations should be revised.

Next, we address the shift and relaxation data.
The relatively short $T_1$ of $^{77}$Se (spin-1/2) in \bise, cf.~Tab.~\ref{tab:data}, 
demands a significant Fermi level density of states (DOS) in the bulk of the material. 
This is supported by our preliminary $^{209}$Bi NMR results, which give  $^{209}T_1 \sim 10$ ms in \bise\, (data not shown), in agreement with Refs. \cite{Young2012,Nisson2013}.
Doping with Cu shortens the $T_1$ of $^{77}$Se by about a factor of 10, likely due to an increase in DOS. 
Hence, we expect the shift in \bise\, and \cubise\, to consist of chemical and Knight shift contributions, the latter being larger for \cubise. We thus interpret the shift differences between both materials by a change in Knight shift. 
For the change in the isotropic Knight shift, i.e., $K_{\rm iso}($\cubise)$-K_{\rm iso}($\bise), we find -17~ppm (\Seo)
 and +99~ppm (\Sei) (cf. Tab.~\ref{tab:data} and note that $K_{\rm iso}=(K_{\parallel}+2K_{\perp})/3$).
Possible contributors to the isotropic Knight shift are the Fermi contact interaction and the core polarization (expected to make a negative contribution because an unpaired electron in the 4p shell gives approximately {\bf-}5 T \cite{Carter1977}). In presence of spin-orbit coupling, orbital effects can also result in a (negative) contribution~\cite{Pavarini2006}. 
For the Knight shift anisotropies, i.e., $\Delta K($\cubise)$-\Delta K($\bise), we find +80 ppm (\Seo) and -210~ppm (\Sei) (with $\Delta K=2(K_{\parallel}-K_{\perp})/3$). 
In order to disentangle the different shift contributions, numerical calculations will be necessary for materials with known carrier concentration.

For the remainder of this work, we argue that the large field-independent linewidths of both resonances are due to indirect internuclear coupling mediated by {\em bulk} electrons.
This mechanism has not been discussed in previous NMR studies of TIs.
Specifically, we argue that the Se linewidths are dominated by the indirect scalar coupling between $^{77}$Se nuclei and the 100\% abundant $^{209}$Bi nuclei (the coupling between $^{77}$Se nuclei can be neglected due to low natural abundance and small spin).
We approximate the indirect scalar coupling between $^{77}$Se and $^{209}$Bi by a Hamiltonian $H=\sum_{i j} J_{i j} {\bf I}_i \cdot {\bf I}_j$ ($i={\rm Se}$,~$j={\rm Bi}$).
Then, the second moment that describes the width of a particular Se resonance line ($i$) is given by~\cite{Abragam1961}
\beq
\label{eq:width01}
\left<\Delta\omega^2\right>_i=\sum_j\frac{I_j(I_j+1)}{3 \hbar^2}{J_{ij}^2},
\eeq
where $I_j=9/2$ is the nuclear spin for $^{209}$Bi.
The exchange coupling constants ($J_{i j}$), which include the contact hyperfine coupling constants, can be calculated in the usual way (see e.g. Ref.~\cite{Ziener2004}).
From this calculation, it becomes apparent that topological insulators such as \bise\, and \bite\, are unusual in that they can mediate a significant Bloembergen-Rowland (BR)~\cite{Bloembergen1955} coupling between nuclear spins, even when the Fermi level is placed inside the bulk gap.
This is in part due to the small bandgaps of these materials (which implies a relatively long range of the BR coupling), and in part due to the strong interband matrix elements of the electronic spin operator~\cite{Yu2010}.

\begin{figure}[t]
\includegraphics[width=0.95\columnwidth]{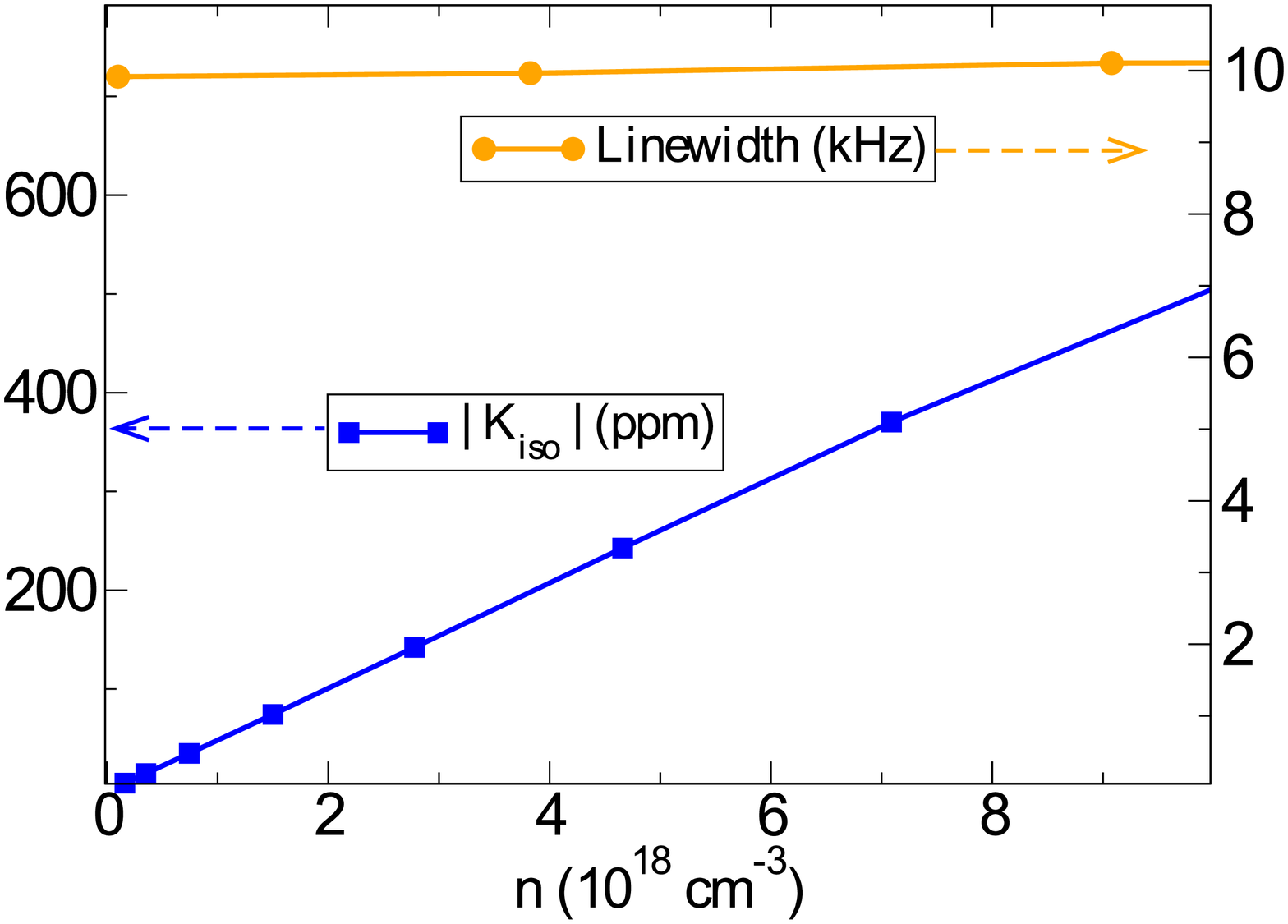}
\caption{\label{fig:kwl}(color online)
Calculated isotropic Knight shift ($|K_{\rm iso}|$) and linewidth ($\sqrt{\langle\Delta\omega^2\rangle}\sqrt{\ln{4}}/\pi$) as a function of the carrier density, for Se$_{\rm in}$.
The model Hamiltonian and the electronic $g$-factors are adopted from Ref.~\cite{Liu2010}.
The value of the contact hyperfine interaction is chosen to yield a Knight shift magnitude in reasonable agreement with experiment.
This same value is then used to calculate $J_{ij}$ and thereafter the linewidth, following Eq.~\eqref{eq:width01} and Ref.~\cite{Ziener2004}. 
}
\end{figure}

In Fig.~\ref{fig:kwl}, we plot  the linewidth calculated with \eqref{eq:width01} for Se$_{\rm in}$, following a model calculation for $J_{ij}$ as a function of carrier concentration.
This theoretical result, which relates the linewidth to the Knight shift, supports the hypothesis that indirect nuclear coupling can play an important role  in the NMR linewidths of TIs.
Moreover, the weak dependence of the linewidth on the carrier concentration reflects the fact that the BR contribution to $J_{i j}$ dominates over the Ruderman-Kittel-Kasuya-Yoshida contribution.

On the experimental side, field-\textit{independent} linewidths in excess of what might be expected from internuclear dipole interaction for the spin-1/2 (non-quadrupolar) $^{77}$Se nuclei point immediately to indirect nuclear spin coupling between $^{77}$Se and $^{209}$Bi.
The largely isotropic linewidths that we measure support this interpretation, with differences between the two sites caused mainly by different hyperfine couplings.
We find from spin echo experiments that ${T_{\rm 2G} \gg T_{\rm 2G}^*}$. 
This confirms a large inhomogeneous broadening of the $^{77}$Se NMR, e.g., as given by \eqref{eq:width01}.  
Fast sample rotation about the magic angle did not result in any significant narrowing or appearance of spinning sidebands for the \bise\, powder (data not shown), as noticed before \cite{Taylor2012}, thereby supporting the explanation in terms of indirect scalar coupling. 

The question arises as to what causes the rather short spin echo decays since conventional homonuclear dipolar coupling is far too weak.  
Note that the ratios between the $T_{\rm 2G}$s for one sample are very similar to the ratios between the $T_{\rm 2G}^*$s (i.e., inverse linewidth ratios, cf.~Tab.~\ref{tab:data}) and hence are determined by the hyperfine couplings. 
The fact that $T_{\rm 2G}$ and $T_{\rm 2G}^*$ are smaller for the less abundant \Sei\, suggests a larger hyperfine coupling for this site (in agreement with shifts and $T_1$).

Given that Cu doping decreases all $T_{\rm 2G}$ by about a factor of 2.4, while the linewidths remain unchanged \footnote{Preliminary field-dependent $^{77}$Se NMR measurements in a Cu$_{0.1}$\bise\, single crystal give for $B_0\parallel c$ slightly smaller field-\textit{independent} linewidths of about 7 kHz (\Seo) and 23 kHz (\Sei). The field-dependent linewidth for \Seo\, was found to be $\sim0.6$ kHz/T, slightly  bigger than in \bise, probably due to distribution of shifts.}, the interaction causing the $^{77}$Se linewidths cannot be responsible for the echo decay.
With a double resonance experiment involving \Seo\, and \Sei\, we have confirmed that the $^{77}$Se-$^{77}$Se coupling contributes only weakly to the spin echo decay \footnote{
While observing the spin echo on \Seo\, with a $\pi$-pulse at $\tau=150~\mu$s, we have inverted also \Sei\, for comparison (right after the regular $\pi$-pulse). We found a 10\% decrease in echo intensity when both Se were flipped, from which we estimate a decay constant due to \Seo-\Sei\, coupling $T_{\rm 2G,out-in}\approx580~\mu$s (an order of magnitude shorter than expected from dipolar coupling between \Seo\, and \Sei). This effect argues in favor of $^{77}$Se-$^{77}$Se indirect coupling, which is expected to contribute only weakly to the spin echo decay of \Seo\, ($T_{\rm 2G,out-out}\geq T_{\rm 2G,out-in}$), since \Seo\, has a smaller hyperfine coupling.}. 

It seems possible that $^{209}$Bi spin flips, e.g., due to spin-lattice relaxation, will induce fluctuations of the $^{77}$Se local field via the $^{77}$Se-$^{209}$Bi indirect coupling, thereby causing spin echo decay. 
If we assume the amplitudes of the fluctuating fields at $^{77}$Se given by the $^{77}$Se-$^{209}$Bi indirect coupling, we can fit our decays to the theory of Recchia {\em et al.}~\cite{Recchia1996}, cf. Fig.~\ref{fig:DecayLWvsB}(a). We obtain a single correlation time for each material (for both Se sites), $2.1\pm0.3$ ms in \bise\, and $150\pm50~\mu$s in \cubise.
The \bise\, correlation time complies with our nonselectively excited $^{209}T_1$ value of 10 ms (the single level lifetime for $I=9/2$ is about a factor of 10 smaller \cite{Slichter2010}). 
A tenfold shorter  $^{209}T_1$ in \cubise\, is expected due to the factor 10 decrease of  $^{77}T_1$ upon Cu doping, cf.~Tab.~\ref{tab:data}. 
We conclude that the $^{209}$Bi level lifetime together with the indirect coupling accounts for the spin echo decays. 

A strong indirect coupling should also affect the $^{209}$Bi NMR. 
Here, the coupling is dominated by $^{209}$Bi-$^{209}$Bi interactions, but does not lead to exchange narrowing \cite{Abragam1961} since the local symmetry at the Bi site causes a sizable ($\sim$140 to 170 kHz)  quadrupole shift of the $^{209}$Bi nuclear levels \cite{Young2012, Nisson2013}, such that nuclear neighbors may not be able to participate in exchange if they are in different spin states. 
Furthermore, the quadrupole shift can vary from one nucleus to the next, due to strain caused by imperfections. 
As a result, the line broadening and spin echo behavior of ${^{209}}$Bi NMR can be quite complicated and may depend on the impurity levels.

We have confirmed, with $^{209}$Bi NMR on our \bise\, single crystal, that all 9 lines from quadrupole splitting have very similar yet large linewidths (\cite{Young2012, Nisson2013}, data not shown). 
This shows that quadrupolar broadening does not dominate the linewidths. 
In preliminary field-dependent measurements on \bise\, powder, we find a large field-\textit{independent} linewidth of about $44\pm4$ kHz for the $^{209}$Bi central transition, in agreement with $46\pm2$ kHz measured in our single crystal at the magic angle and 9.4 Tesla.
This width strongly exceeds the estimated dipolar linewidth of $\sim$1.5 kHz, and hence must be caused by indirect coupling.

Unusually large ${^{209}}$Bi linewidths have been noticed in previous work \cite{Young2012, Nisson2013, Nisson2014, Mukhopadhyay2015}, but indirect coupling was not invoked as a possible origin.  
Besides, rather fast ${^{209}}$Bi NMR spin echo decays were observed, in particular in the central region of the spectra, but could not be explained. 
Interestingly, the linewidth from Ref.~\cite{Nisson2013} increases with decreasing carrier concentration (larger linewidths for more homogeneous samples grown with excess Se).
Most of these results appear compatible with a large indirect coupling between the $^{209}$Bi nuclei in the presence of quadrupole interaction.

To conclude, we have presented a detailed $^{77}$Se NMR study of \bise\, and \cubise. 
First, we have identified and characterized the two resonances from \Sei\, and \Seo. 
Secondly, we have measured large, field-independent NMR linewidths that suggest a strong Bloembergen-Rowland internuclear coupling mediated by bulk electrons. 
Our study may have fundamental implications for future NMR experiments in other topological materials, and may pave the way for a NMR-based characterization of topological surface states.

$~$
\begin{acknowledgments}
We acknowledge the help of F. H\" ofer and financial support from the University of Leipzig and the DFG within the Graduate School BuildMoNa. 
J.H. and N.G. thank B. Rosenow and B. Zocher for helpful discussions.
I.G. thanks Qu\'ebec's RQMP and Canada's NSERC for funding, Calcul Qu\'ebec and Compute Canada for computer resources, and J. Quilliam for discussions.
 
\end{acknowledgments}

\vspace{1cm}

\bibliography{TopIns}
\bibliographystyle{apsrev4-1}

\end{document}